# Application of Time-Fractional Order Bloch Equation in Magnetic Resonance Fingerprinting

*Haifeng Wang[1], Lixian Zou[1,2], Huihui Ye[3], Shi Su[1], Yuchou Chang[4], Xin Liu[1], Dong Liang[1] ***

[1]Paul C. Lauterbur Research Center for Biomedical Imaging,
Shenzhen Institutes of Advanced Technology, Chinese Academy of Sciences
[2]Shenzhen College of Advanced Technology, University of Chinese Academy of Sciences
[3]State Key Laboratory of Modern Optical Instrumentation,
College of Optical Science and Engineering, Zhejiang University
[4]Department of of Computer Science and Engineering Technology,
University of Houston-Downtown

## ABSTRACT

Magnetic resonance fingerprinting (MRF) is one novel fast quantitative imaging framework for simultaneous quantification of multiple parameters with pseudo - randomized acquisition patterns. The accuracy of the resulting multi-parameters is very important for clinical applications. In this paper, we derived signal evolutions from the anomalous relaxation using a fractional calculus. More specifically, we utilized time-fractional order extension of the Bloch equations to generate dictionary to provide more complex system descriptions for MRF applications. The representative results of phantom experiments demonstrated the good accuracy performance when applying the time-fractional order Bloch equations to generate dictionary entries in the MRF framework. The utility of the proposed method is also validated by in-vivo study.

## 1. INTRODUCTION

Magnetic resonance fingerprinting (MRF) is a time-efficient acquisition and reconstruction framework to provide simultaneous measurements of multiple parameters including the $T_1$ and $T_2$ maps [1]. Instead of acquiring the steady state signal with constant sequence parameter - as done in conventional techniques[2], MRF generates the temporal and spatial incoherent signal evolutions by using *pseudo-randomized acquisition patterns, such as varying flip angles (FAs), repetition times (TRs) and variable density spiral trajectories. After acquisition, multi-parametric maps were reconstructed through a pattern recognition algorithm. More specifically, a series of spiral interleaves was first transformed into high aliased image series. Then, a pixel-wise matching was developed to match the signal evolutions to a set of pre-computed dictionary entries. The entries were theoretical signal evolutions simulated by the Bloch equation or extended phase graph (EPG) [3] with the various $T_1$ and $T_2$ combinations. Dictionary generation is one of the most important roles in MRF, which, however, is time consuming with a wide combination of $T_1$, $T_2$ and off-resonance frequency [1, 4]. Fortunately, dictionary needs to be generated only once with specific sequence. Moreover, various methods to optimize time efficiency in MRF have been proposed in recent studies [5-7]. Since the pattern recognition nature in MRF, it is more robust to various image artifacts like motion and under-sampling.

A challenge in MRF is the accuracy of the resulting multi-parameters. Many groups have shown $T_1$ and $T_2$ deviation occurred even after correction [8, 9]. Since MRI is a more complex system, there are many anomalous cases being observed such as stretched-exponential or power-law behavior [10, 11]. Fractional order generalization of the Bloch equations, thus, is more flexible to describe the dynamics of complex phenomenon including the anomalous NMR relaxation phenomenon. Actually, the fractional order model has shown to be useful for describing dielectric and viscoelastic relaxation in complex, heterogeneous materials.

In this paper, we proposed one novel method to apply the time-fractional order extension of the Bloch equations into MRF data matching. Firstly, we described the development of a time-fractional order model for NMR or MRI relaxation, which leads to Mittag-Leffler function [10] for time domain longitudinal (or spin-lattice) and transverse (or spin-spin) relaxation. Then, the conventional first order model of the Bloch equations in the MRF framework was alternated by the time-fractional order model to generate dictionary entries in the MRF framework. We validated the effectiveness of the proposed method using phantoms and in vivo brain experimental data and its performance was measured by quantitatively comparing the results with those obtained from gold standard methods. These representative results illustrated that the proposed approach has better accuracy performance than the conventional approach. The preliminary results have been published previously in the Ref. [12].

* This work is supported in part by the National Natural Science Foundation of China (61871373, 81729003) and the Natural Science Foundation of Guangdong Province (2018A0303130132).

## 2. PROPOSED METHOD

### 2.1. Classical Bloch Equation

The Bloch equation is a phenomenological equation that can be written as:

$$\frac{dM_z(t)}{dt} = -\frac{M_z(t) - M_0}{T_1}, \quad (1a)$$

$$\frac{dM_{xy}(t)}{dt} = -i\omega_0 M_{xy}(t) - \frac{M_{xy}(t)}{T_2}, \quad (1b)$$

where $M_z(t)$ and $M_{xy}(t)$ denote time-varying nuclear magnetization over a unit volume in longitudinal and transverse direction. $M_0$ is the equilibrium magnetization, and, $T_1$ and $T_2$ is spin-lattice and spin-spin relaxation times, respectively. The left-hand sides prescribe the time derivatives of the nuclear magnetization at a static magnetic field. The right-hand sides describe precession the return of the longitudinal component of the magnetization to equilibrium and the decay of the transverse component, respectively.

### 2.2. Time-fractional Order Bloch Equation

Time-fractional operators applied in Bloch equations were proposed by an extensive number of researchers [10-12]. Magin's fractionalizing approach [10] to incorporate the Caputo derivative into the left side of the Bloch equations in Eq. (1a) and Eq. (1b) will be adopted in this work:

$$^C_0D_t^\alpha M_z(t) = -\frac{M_z(t) - M_0}{T'_1}, \quad (2a)$$

$$^C_0D_t^\beta M_{xy}(t) = -i\bar{\omega}_0 M_{xy}(t) - \frac{M_{xy}(t)}{T'_2}, \quad (2b)$$

Where $\alpha$ and $\beta$ represent the time-fractional order of anomalous relaxation of $T_1$ and $T_2$. $\bar{\omega}_0 = \omega_0/\tau_2^{\beta-1}$ has the unit of $(s)^{-\beta}$, $T'_1 = \tau_1^{\alpha-1}T_1$ and $T'_2 = \tau_2^{\beta-1}T_2$ have the units of $(s)^\beta$. $\tau_1$ and $\tau_2$ are fractional time constants to maintain a consistent set of units on both sides of Eq. (2). $^C_0D_t^*$ denotes the Caputo derivative operator with order * in time. * represents $\alpha$ or $\beta$. The definition and properties of the fractional derivative are given according to Ref. [10-12].

The solutions to Eq. (2a) and Eq. (2b) can be solved using fractional calculus or the Laplace transformation in a reference framework rotating with $\omega_0$ as

$$M_z(t) = M_z(0) E_\alpha\left(\frac{-t^\alpha}{T'_1}\right) + \frac{M_0}{T'_1} t^\alpha E_{\alpha,\alpha+1}\left(\frac{-t^\alpha}{T'_1}\right), \quad (3a)$$

$$M_{xy}(t) = M_{xy}(0) E_\beta\left(\frac{-t^\beta}{T'_2}\right), \quad (3b)$$

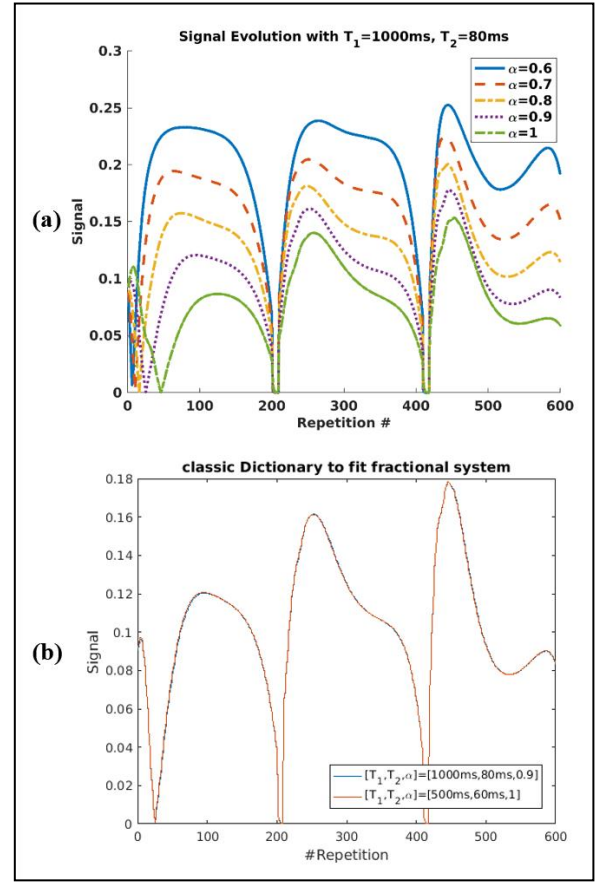

**Figure 1.** (a) Illustration of signal evolutions with $T_1$=1000ms and $T_2$=80ms when α sets to increase 0.6 to 1 with step 0.1 (supposed α=β for example). (b) Using conventional dictionary to match the supposed complex environment in a voxel with $T_1$=1000ms and $T_2$=80ms, and the complexity of environment performed as α=β=0.9.

Where $E_\alpha(t)$ and $E_{\alpha,\alpha+1}(t)$ are the single and two-parameter Mittag-Leffler function, respectively. In the case of α=β=1, the Mittag-Leffler functions corresponds to the classical Bloch equations, leading to the conventional exponential relaxation processes. Note that time-fractional order Bloch equations can be defined separately to describe the longitudinal and transvers anomalous relaxations.

## 3. RESULTS

In this section, we firstly generated signal evolutions using both classical and time-fractional order Bloch equations. $T_1$ and $T_2$ relaxation values were set as 1000ms and 80ms, while α and β were set to range from 0.6 to 1 with step 0.1. FAs were varied as sinusoidal at each repetition of the pattern to smooth transient state of the magnetization varying, while the max FAs were set to same as Ref. [4]. Fig.1 (a) shows the variance of signal evolutions with a set of fractional orders. The process in the complex dynamic

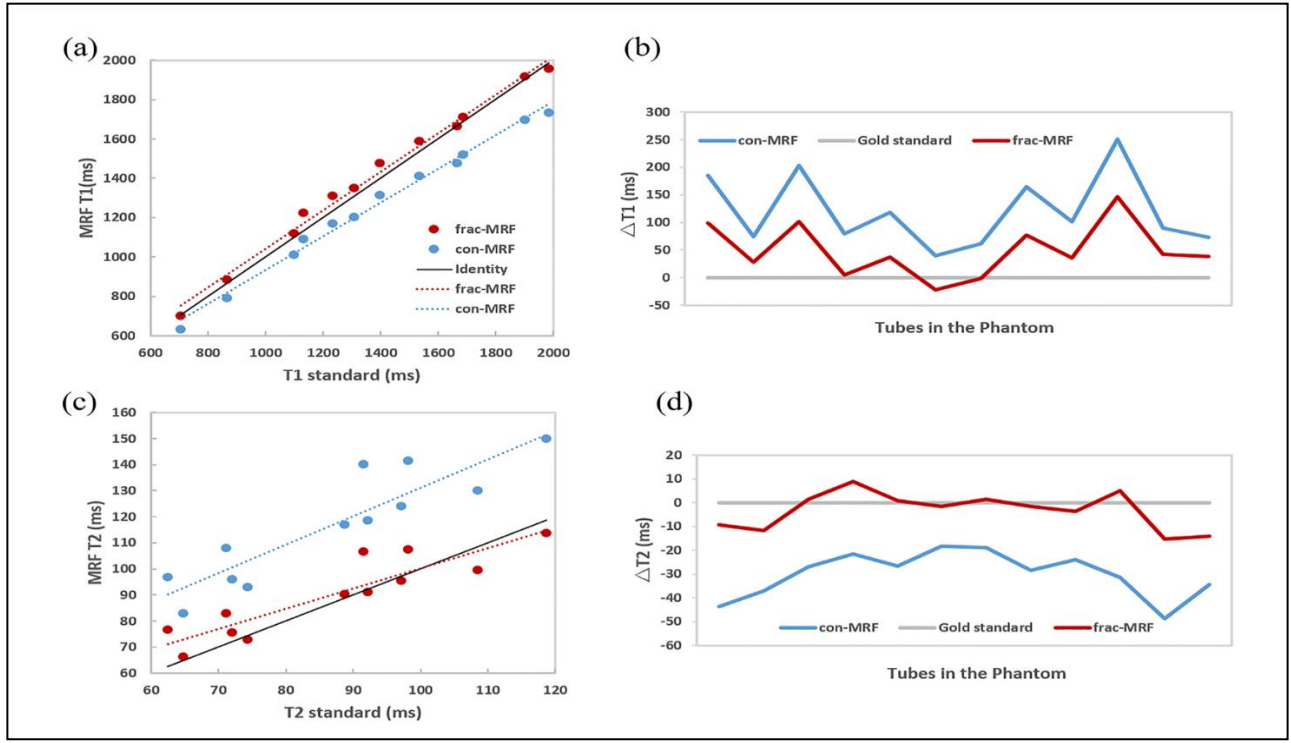

**Figure 2.** Left: Accuracy of the phantom scan: $T_1$ (a) and $T_2$ (c) values from 12 tubes from the phantom using conventional dictionary (Con-MRF) and the proposed dictionary (Frac-MRF). Right: Performance comparison: the plots of bias for $T_1$ (b) and $T_2$ (d) in 12 tubes from the phantom using conventional and time-fractional Bloch equations to generate dictionary.

system ( when α=β=0.9 ) in Fig. 1(b) is interpreted to underestimate the $T_1$ and $T_2$ values when using the conventional dictionary

In phantom study, dictionary entries used for MRF matching were generated using the two models mentioned above. The dictionary was generated for a wide range of possible $T_1$ values (range from 100 to 4500 ms), $T_2$ values (range from 10 to 1000 ms), α and β values (increase from 0.96 to 1.1 with step 0.01). The MRF image series (full sampled with 600 time points) of the phantom (12 tubes; mixtures of Agar and $MnCl_2$) were acquired on a commercial 3 Tesla Prisma scanner (Siemens Healthcare, Erlangen, Germany) with a 16-channel head coils. The resolution of the images was 1×1 $mm^2$ in a field of view (FOV) 220×220 $mm^2$. The resulting $T_1$ and $T_2$ values were compared to the standard values estimated using conventional spin echo sequence. α=0.98 and β=1.08 were selected as the best fractional order for the time-fractional order model to approach $T_1$ and $T_2$ standards. Accuracy and bias of two dictionary models have been shown in Fig. 2.

In the in-vivo study, a high under-sample variable-density spiral trajectory was used for data acquisition. To reduce the trajectory artefacts, sliding window [6] was used in our study. We firstly get a series of highly under sample images, then get a better images course with the sliding window size of 36 shots. The space resolution for the parametric maps was 1×1 $mm^2$ in a field of view (FOV) 300×300 $mm^2$ with matrix size of 320×320. The proposed protocols were approved by our Institutional Reviews Board (IRB), and written informed consent was obtained from the subject prior to perform scanning. Parameter maps obtained using the conventional MRF (Con-MRF) and time-fractional order MRF (Frac-MRF) model have been shown in Fig.3.

## 4. DISCUSSION AND CONCLUSION

The process in the complex dynamic system demonstrates similar signal evolution with conventional relaxation. However, it would be interpreted to underestimate $T_1$ and $T_2$ values by the conventional relaxation model in simulation. The accuracy of $T_1$ and $T_2$ values is improved perfectly using Frac-MRF in phantom experiment. It also improves the $T_2$ value accuracy compared to Con-MRF, but a deviation is shown for the scattered values. $T_1$ and $T_2$ values deviation from the standard values become severe when tubes in phantom are off the magnetic field center, which can be corrected by simulating the $B_1$ into the dictionary. The $T_1$ and $T_2$ values in gray and white matter are smaller than the previously published ones generated with spin echo–based mapping techniques when using conventional MRF framework [9, 13]. The proposed fractional method shows its ability to approach the parameter values in MRF to the spin echo–based ones as shown in Ref.[13].

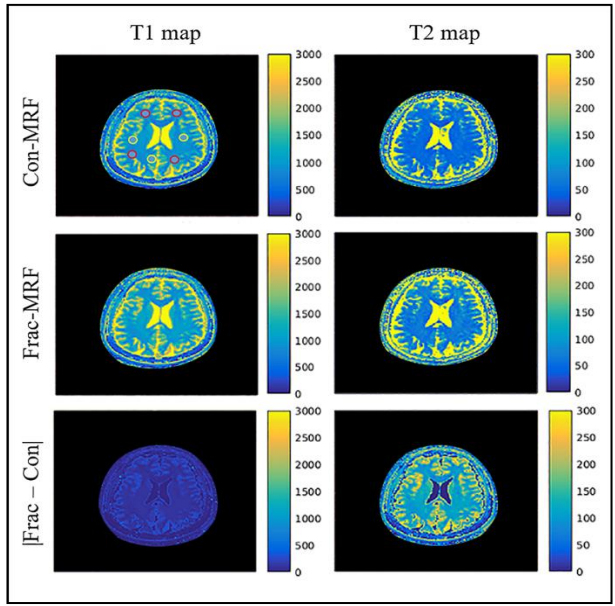

**Figure 3.** Parameter maps from Frac-MRF and Con-MRF, and the difference maps between them. Left column: $T_1$ maps, Right column: $T_2$ maps; Top to bottom: Maps from Con-MRF, Frac-MRF, and difference between Frac-MRF and Con-MRF. $T_1$ and $T_2$ values of white matter (red circle) in Con-MRF are 827±52 ms and 52±8 ms, while in Frac-MRF are 924±59 ms and 70±12 ms. $T_1$ and $T_2$ values of gray matter (yellow circle) in Con-MRF are 1398±118 ms and 67±5 ms, while in Frac-MRF are 1571±134 ms and 93±7 ms. $T_1$ and $T_2$ values are high correlated to the literature ones [13] using the proposed method.

In presented work, we utilize the simplest MRF framework to illustrate the application of time-fractional order dictionary in MRF. α and β are considered as global parameters to regulate the system as well as they are manually selected, which might lead to a coarse result. Even so, the representative results demonstrate Frac-MRF is able to approach to the standard values better than the one from the conventional exponential relaxation model. Since there are various kinds of definitions for fractional derivatives, it would be interesting to study more complex dynamic NMR system in considering time delay, multi-term and transient chaos in applications of fractional calculus [14].

In sum, we proposed a dictionary generation model for MRF by using Mittag-Leffler functions derived from the time-fractional order generalization of Bloch equations. This time-fractional model is particularly useful to describe anomalous NMR relaxation phenomenon due to nonlocal interactions and system memory in complex, heterogeneous materials, such as biological tissue. The advantages offered by the proposed method have been demonstrated in terms of its general extension of the classical simple exponential relaxation, which presents an opportunity to accurately describe a wider range of complex situations in MR system.

## 5. REFERENCES

[1] D. Ma, V. Gulani, N. Seiberlich, K. Liu, J. L. Sunshine, J. L. Duerk, and M. A. Griswold, "Magnetic Resonance Fingerprinting," *Nature*, vol. 495, pp.187-192, 2013.
[2] H.L. Cheng, N. Stikov, N.R. Ghugre, G.A. Wright, "Practical medical applications of quantitative MR relaxometry," *J Magn. Reson. Imaging*, vol. 36, pp. 805–824, 2012.
[3] M. Weigel, "Extended phase graphs: dephasing, RF pulses, and echoes—pure and simple," *J Magn. Reson. Imaging*, vol. 41, pp. 266–295, 2015.
[4] Y. Jiang, D. Ma, N. Seiberlich, V. Gulani, and M. A. Griswold, "MR fingerprinting using fast imaging with steady state precession (FISP) with spiral readout," *Magn. Reson. Med.*, vol. 74, pp. 1621–1631, 2015.
[5] H. Ye, D. Ma, Y. Jiang, S. F. Cauley, Y. Du, L. L. Wald, M. A. Griswold, and K. Setsompop, "Accelerating magnetic resonance fingerprinting (MRF) using t‐blipped simultaneous multislice (SMS) acquisition," *Magn. Reson. Med.*, vol. 75, pp. 2078-2085, 2016.
[6] X. Cao, C. Liao, Z. Wang, Y. Chen, H. Ye, H. He, and J. Zhong, "Robust sliding-window reconstruction for Accelerating the acquisition of MR fingerprinting," *Magn. Reson. Med.*, vol. 78, pp. 1579-1588, 2017.
[7] B. Zhao, K. Setsompop, E. Adalsteinsson, B. Gagoski, H. Ye, D. Ma, Y. Jiang, P. E. Grant, M. A. Griswold, and L. L. Wald, "Improved magnetic resonance fingerprinting reconstruction with low-rank and subspace modeling: A Subspace Approach to Improved MRF Reconstruction". *Magn. Reson. Med.*, vol. 79, pp. 933-942, 2018.
[8] J. Assländer, S.J. Glaser, J. Hennig, "Pseudo steady-state free precession for MR-fingerprinting," *Magn. Reson. Med.*, vol. 77, pp. 1151–1161, 2017.
[9] G. Körzdörfer, Y. Jiang, P. Speier, J. Pang, D. Ma, J. Pfeuffer, B. Hensel, V. Gulani, M. Griswold, and M. Nittka. "Magnetic resonance field fingerprinting," Magn. Reson. Med., vol. 00, pp. 1–13, 2018.
[10] R. L. Magin, Weiguo Li, M. P. Velasco, J. Trujillo, D. A. Reiter, A. Morgenstern, R. G. Spencer, "Anomalous NMR relaxation in cartilage matrix components and native cartilage: Fractional-order models", *Journal of Magnetic Resonance*, vol. 210, pp. 184-191, 2011.
[11] S. Qin, F. Liu, I. W. Turner, Q. Yu, Q. Yang, and V. Vegh, "Characterization of anomalous relaxation using the time-fractional Bloch equation and multiple echo T2*-weighted magnetic resonance imaging at 7 T," *Magn. Reson. Med.*, vol. 77, pp. 1485-1494, 2017.
[12] H. Wang, L. Ying, X. Liu, H. Zheng, D. Liang. "MRF-FrM: A Preliminary Study on Improving Magnetic Resonance Fingerprinting Using Fractional-order Models". *Proc. 26th Annual Meeting of ISMRM*, Paris, France, 2018.
[13] J. Bojorquez, S. Bricq, C. Acquitter, F. Brunotte, P. M. Walker, and A. Lalande, "What are normal relaxation times of tissues at 3 T?" *Magnetic resonance imaging*, vol. 35, pp. 69-80, 2017.
[14] S. Bhalekar, V. Daftardar-Gejji, D. Baleanu, R. L. Magin, "Fractional Bloch equation with delay," *Computers & Mathematics with Applications*, vol. 61, pp. 1355-1365, 2011.